\documentclass[conference]{IEEEtran}
\IEEEoverridecommandlockouts

\usepackage{cite}
\usepackage{amsmath,amssymb,amsfonts, booktabs, tabularx, multicol}
\usepackage{algorithmic}
\usepackage{graphicx}
\usepackage{textcomp}
\usepackage{xcolor}

\newcolumntype{C}{>{\centering\arraybackslash}X}
\newcolumntype{R}{>{\raggedleft\arraybackslash}X}
\newcolumntype{L}{>{\raggedright\arraybackslash}X}

\newcommand{\ER}{$\text{ER}_{\le 20^\circ}$}
\newcommand{\Fone}{$\text{F}_{\le 20^\circ}$}
\newcommand{\LE}{$\text{LE}_\text{CD}$}
\newcommand{\LR}{$\text{LR}_\text{CD}$}
\newcommand{\ESELD}{$\mathcal{E}_{\text{SELD}}$}

\newcommand{\cmmnt}[1]{\ignorespaces} 
\makeatletter
\newcommand{\linebreakand}{%
  \end{@IEEEauthorhalign}
  \hfill\mbox{}\par
  \mbox{}\hfill\begin{@IEEEauthorhalign}
}
\makeatother

\def\BibTeX{{\rm B\kern-.05em{\sc i\kern-.025em b}\kern-.08em
    T\kern-.1667em\lower.7ex\hbox{E}\kern-.125emX}}
\begin{document}

\title{Real-Time Sound Event Localization and Detection: Deployment Challenges on Edge Devices\\

\thanks{This research is supported by the Singapore Ministry of Education, Academic Research Fund Tier 2, under research grant MOE-T2EP20221-0014.}
}

\author{

\IEEEauthorblockN{Jun Wei Yeow}
\IEEEauthorblockA{\textit{Smart Nation Trans Lab} \\
\textit{Nanyang Technological University}\\
Singapore \\
junwei004@e.ntu.edu.sg}

\and

\IEEEauthorblockN{Ee-Leng Tan}
\IEEEauthorblockA{\textit{Smart Nation Trans Lab} \\
\textit{Nanyang Technological University}\\
Singapore \\
etanel@ntu.edu.sg}

\and

\IEEEauthorblockN{Jisheng Bai}
\IEEEauthorblockA{\textit{School of Marine Science and Technology} \\
\textit{Northwestern Polytechnical University}\\
Xi'an, China\\ 
baijs@mail.nwpu.edu.cn}

\linebreakand

\IEEEauthorblockN{Santi Peksi}
\IEEEauthorblockA{\textit{Smart Nation Trans Lab} \\
\textit{Nanyang Technological University}\\
Singapore \\
speksi@ntu.edu.sg}

\and

\IEEEauthorblockN{Woon-Seng Gan}
\IEEEauthorblockA{\textit{Smart Nation Trans Lab} \\
\textit{Nanyang Technological University}\\
Singapore \\
ewsgan@ntu.edu.sg}

}

\maketitle

\begin{abstract}
Sound event localization and detection (SELD) is critical for various real-world applications, including smart monitoring and Internet of Things (IoT) systems. Although deep neural networks (DNNs) represent the state-of-the-art approach for SELD, their significant computational complexity and model sizes present challenges for deployment on resource-constrained edge devices, especially under real-time conditions. Despite the growing need for real-time SELD, research in this area remains limited. In this paper, we investigate the unique challenges of deploying SELD systems for real-world, real-time applications by performing extensive experiments on a commercially available Raspberry Pi 3 edge device. Our findings reveal two critical, often overlooked considerations: the high computational cost of feature extraction and the performance degradation associated with low-latency, real-time inference. This paper provides valuable insights and considerations for future work toward developing more efficient and robust real-time SELD systems\footnote{The code to run the inference pipeline can be found here: https://github.com/itsjunwei/Realtime-SELD-Edge}.
\end{abstract}

\begin{IEEEkeywords}
Sound event localization and detection, IoT, real-time inference, direction-of-arrival estimation
\end{IEEEkeywords}

\section{Introduction}
\label{sec:intro}

Sound Event Localization and Detection (SELD) integrates Sound Event Detection (SED) with Direction-of-Arrival (DOA) estimation, enabling systems to identify the onset, offset, and spatial trajectory of sound events. This capability is critical for applications such as autonomous driving \cite{mohmmad2024parametric}, robotic navigation \cite{nakamura2012real_doarobot}, and smart monitoring systems within Internet of Things (IoT)\cite{zhang2024sound, shah2018audio_iot, tan2021extracting}. With growing interest in real-time applications, such as wearable devices \cite{nagatomo2022wearable, yasuda20246dof}, there is an increasing demand for SELD systems that can operate with low latency while maintaining high performance.

State-of-the-art SELD systems often leverage complex deep neural networks (DNNs), such as convolutional recurrent neural networks (CRNNs) \cite{politis2020overview}. However, their substantial computational requirements and large parameter counts pose significant deployment challenges on resource-constrained edge devices often used in real-world scenarios \cite{stefani2022comparison, wyatt2021environmental}. While existing research has focused on improving SELD models for offline performance, there is a notable gap in investigating the unique constraints of real-time, on-device inference, particularly concerning computational efficiency and latency.

This paper explores the challenges of deploying SELD systems for real-time inference on edge devices. We analyze the trade-offs between computational cost and performance across various input features and DNN architectures trained for polyphonic SELD. From this analysis, two key challenges are highlighted: the high computational burden of feature extraction and the performance degradation observed in low-latency, real-time scenarios. These challenges underscore the trade-offs and need for careful design considerations when developing SELD systems optimized for edge deployment. 

The key contributions of this paper are as follows:
\begin{enumerate}
     \item Comprehensive experiments on deploying SELD systems on edge devices to highlight practical challenges associated with real-time inference.
    \item Insights into the trade-offs between computational cost, feature extraction efficiency, and SELD performance for real-time SELD inference systems.
\end{enumerate}

\subsection{Related Work}
\label{section:related_work}

Lightweight DNN-based systems have been extensively explored in related acoustic tasks such as acoustic scene classification \cite{martin2021low, martin2022low} and music source separation \cite{venkatesh_realtime_MSS}. However, efficient SELD systems, particularly for real-time applications, remain underexplored. Brignone et al.\cite{brignone2022efficient} and Perez et al. \cite{perez2020papafil} proposed efficient SELD architectures for ambisonic signals utilizing Quaternion processing and spatial parametric analysis, respectively. Nguyen et al. \cite{Nguyen2022_SALSALite} introduced SALSA-Lite, a computationally efficient feature set for A-format signals, later also utilized for SELD on wearables \cite{yasuda20246dof}. However, these work were all primarily validated in offline settings, and their applicability to real-time inference remains unexplored. 

Research specifically targeting real-time SELD systems is limited, with most existing work focusing on limited sound classes, such as speech \cite{pertila2021mobile} or impulse noise \cite{naing2019real}, which reduces task complexity, but limits generalizability to broader SELD applications. Other studies have concentrated on real-time SED \cite{chavdar2023scarrie, cerutti2020compact, baelde2019real} or DOA estimation \cite{nakamura2012real_doarobot, pavlidi2013real, zhao2013realtime_ssl} independently. Tan et al. \cite{tan2021extracting} deployed a real-time SELD-based monitoring system on an edge device capable of classifying up to 11 sound events, using a hybrid approach that combines DNN-based SED with traditional signal processing for DOA estimation, which may not fully leverage the potential of DNN-based architectures. As such, the goal of this work is to, ultimately, bridge research gaps by exploring the deployment challenges of SELD systems for real-time inference on resource-constrained edge devices.

\section{Analysis method}
\label{section:analysis_method}

The following section covers the setup used to investigate the challenges surrounding real-time SELD inference.

\subsection{Real-time SELD Inference Pipeline}
\label{section:real_time_SELD_pipeline}

\begin{figure}[t]
    \centering
    \includegraphics[width=0.95\columnwidth]{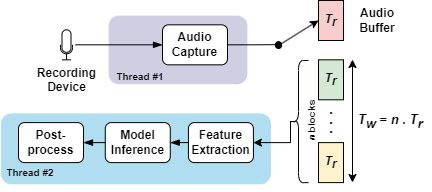}
    \caption{Block diagram of the multi-threaded real-time SELD inference system.}
    \label{fig:rtinferencesystem}
\end{figure}

Figure \ref{fig:rtinferencesystem} illustrates the block diagram of a basic SELD pipeline designed for real-time inference on the edge. The system records audio in blocks of $T_r$ seconds, storing them in an audio buffer. Concurrently, the inference model will process the most recent $n$ consecutive audio blocks from the buffer, totaling $T_w$ seconds. System latency comprises two main components: hardware latency ($T_r$), representing the time taken to record audio, and processing latency, defined as the time required to extract features, perform model inference and handle any post-processing tasks such as tracking \cite{evers2020locata}. Critically, processing latency must not exceed $T_r$ -- the inference pipeline must complete before the next audio block is captured\cite{cerutti2020compact}. All following deployment experiments are conducted on a Raspberry Pi 3 with 1GB RAM\footnote{This pipeline is adaptable and not restricted to the Raspberry Pi 3; it can be implemented on any generic edge device.}.

\subsection{Dataset}
\label{section:dataset}

For experiments, the STARSS23 dataset \cite{shimada2024starss23} is used, which is the dataset for Task 3 of the Detection and Classification of Sound Scenes and Events (DCASE) Challenge 2023. The dataset is recorded in two 4-channel 3-dimensional recording formats: the first-order ambisonics (FOA) and tetrahedral microphone array (MIC) formats. For this work, only the MIC format is considered as the A-format signals lend themselves more flexible to non-tetrahedral microphone array configurations such as for future smart device usage \cite{nagatomo2022wearable, yasuda20246dof}. The STARSS23 dataset consists of real audio recordings in different indoor rooms totaling around 5.5 hours, with 13 sound event classes. In addition, 20 hours of additional synthetic training data\footnote{https://zenodo.org/records/6406873}, provided by the DCASE Challenge organizers, are also added into the training set.

\subsection{SELD Input Features}
\label{section:seld_input_features}

In this work, commonly used input features for the MIC audio format are compared\cite{Nguyen2022_SALSA, cao2019polyphonic, Nguyen2022_SALSALite}. MelGCC is arguably the most popular input feature set, comprising of multichannel Mel spectrograms stacked with generalized cross-correlation with phase transform (GCC-PHAT). Let $X_i(t,f) \in \mathbb{C}$ represent the short-time Fourier transform (STFT) of the $i^\text{th}$ microphone signal, where $t$ and $f$ are the time and frequency indices, respectively. The Mel spectrogram for the $i^\text{th}$ microphone channel is computed by
\begin{equation}
    \label{eqn:melspec}
    \textsc{MelSpec}_i(t,k) = \text{log}
    \left(
    |X_i(t,f)|^2 \cdot
    \mathbf{W}_\text{mel}(f,k)
    \right),
\end{equation}

\noindent where $\mathbf{W}_\text{mel}(f,k)$ denotes the Mel filter for the $k$-th Mel index. The GCC-PHAT feature is calculated for each time frame $t$ and each microphone pair ($i,j$) as follows \cite{cao2019polyphonic}:

\begin{equation}
\label{eqn:gccphat}
    \text{GCC-PHAT}_{i,j}(t, \tau) = \mathcal{F}^{-1}_{f\rightarrow\tau}\left[
    \frac{X_i(t,f)X^H_j(t,f)}{|X_i(t,f)X^H_j(t,f)|}
    \right],
\end{equation}

\noindent where $\tau$ is the time lag, constrained by $|\tau| \leq f_sd_\text{max}/c$, $\mathcal{F}^{-1}$ denotes the inverse Fourier transform and $(\cdot)^H$ is the Hermitian transpose; $f_s$ is the sampling rate, $c=343$m/s is the speed of sound, and $d_\text{max}$ is the largest distance between any two microphones. When GCC-PHAT features are stacked with Mel spectrograms, the range of $\tau$ is truncated to $(-K/2, K/2]$, where $K$ corresponds to the number of Mel bands. Another popular set of input features is SALSA-Lite \cite{Nguyen2022_SALSALite} which consists of linear scale spectrograms stacked with normalized interchannel phase differences (NIPD), using the first channel as reference, calculated by

\begin{equation}
\label{eqn:nipd}
    \text{NIPD}(t,f) = 
    -\frac{c}{2 \pi f}
    \text{arg}(X_{2:4}(t,f)X_1^*(t,f))
    \in \mathbb{R}^{M-1}.
\end{equation}

Using the Mel frequency scale can reduce the dimensionality of frequency indices in the input features, thereby decreasing the computational load during inference \cite{zhao2013realtime_ssl}. Therefore, we also experiment with applying the Mel frequency scale to the SALSA-Lite features, termed SALSA-Mel. The Mel scale NIPDs are calculated by\cite{huang2023dynamic_salsamel}

\begin{equation} 
    \label{eqn:salsa_mel}
    \textsc{MelNIPD}(t,k) = 
    \left(
    \text{NIPD}(t,f) \cdot
    \mathbf{W}_\text{mel}(f,k)
    \right).
\end{equation}

All feature extraction is performed at a sampling rate of 24 kHz using 512-point FFT, with Hann window of length 512 samples and a hop size of 300 samples. For the features in the Mel frequency scale, $K=128$ Mel bands are used. For SALSA-Lite, the frequency bin cutoff specified in \cite{Nguyen2022_SALSALite} is used, resulting in 191 frequency bins.

\subsection{Model Training}

All models investigated adopt the multi-activity Cartesian coordinate direction-of-arrival (multi-ACCDOA) vector output format\cite{shimada_multiaccdoa}, utilizing the auxiliary duplicating permutation invariant training framework with the mean squared loss as the loss function. Models were trained for a total of 100 epochs with a batch size of 32 using the Adam optimizer, starting with an initial learning rate of $1\times10^{-3}$, which linearly decreases to $1\times10^{-5}$ during the final $30\%$ of epochs. Frequency Shifting augmentation \cite{Nguyen2022_SALSA} is applied on-the-fly during training.

\subsection{SELD Evaluation Metrics}
\label{section:seld_eval_metrics}

The official SELD metrics from the DCASE 2023 Challenge Task 3 are used to evaluate SELD performance \cite{politis2020overview}. These metrics include the macro-averaged location-dependent error rate (\ER) and F-score (\Fone) for SED, and the class-dependent localization error (\LE) and localization recall (\LR) for DOA, and an aggregated SELD error (\ESELD) calculated by

\begin{equation}
    \label{eqn:seld_error}
    \mathcal{E}_\text{SELD} = 
    \frac{\left(
    \text{ER}_{\le20^\circ} + 
    (1 - \text{F1}_{\le20^\circ}) +
    \dfrac{\text{LE}_\text{CD}}{180^\circ} +
    (1 - \text{LR}_\text{CD})    
    \right)}
    {4}.
\end{equation}

\noindent An effective SELD system should have low $\text{ER}_{\le20^\circ}$, $\text{LE}_\text{CD}$, and \ESELD, and high $\text{F1}_{\le20^\circ}$ and $\text{LR}_\text{CD}$.

\section{Experimental Results and Discussion}
\label{section:experiment_results_discussion}

\subsection{Impact of Shorter Input Window Durations}
\label{section:offline_seld_inference}

\begin{table}[t]
    \centering
    \caption{Performance of various SELD models and input feature sets with decreasing $T_w$. The best metrics for each variation are highlighted in bold, while the worst metrics are underlined.}
    \begin{tabularx}{\columnwidth}{lc rrrrr}
    \toprule
        System & $T_w$ (s) & \ER & \Fone & \LE & \LR & \ESELD \\
        \midrule

        \multicolumn{2}{l}{\bfseries \textsc{SALSA-Lite}}\\
        
        \midrule

        SELDNet 
        & 5 & \textbf{0.645} & 0.258 & $\textbf{24.7}^\circ$ & 0.436 & \textbf{0.511} \\
        & 4 & 0.650 & \textbf{0.267} & $27.1^\circ$ & 0.438 & 0.524 \\
        & 3 & \underline{0.686} & 0.231 & $28.4^\circ$ & 0.444 & 0.542 \\
        & 2 & 0.659 & 0.247 & $29.4^\circ$ & \textbf{0.473} & 0.526 \\
        & 1 & 0.673 & \underline{0.223} & $\underline{31.3}^\circ$ & \underline{0.421} & \underline{0.551} \\

        \midrule

        ResNet-18 
        & 5 & \textbf{0.593} & \textbf{0.260} & $29.6^\circ$ & 0.432 & \textbf{0.516} \\
        & 4 & 0.624 & 0.259 & $\textbf{25.4}^\circ$ & \textbf{0.442} & 0.521 \\
        & 3 & 0.635 & 0.259 & $\underline{30.4}^\circ$ & \textbf{0.442} & 0.526 \\
        & 2 & 0.624 & 0.254 & $27.7^\circ$ & 0.425 & 0.525 \\
        & 1 & \underline{0.654} & \underline{0.246} & $27.5^\circ$ & \underline{0.414} & \underline{0.537} \\

        \midrule

        \multicolumn{2}{l}{\bfseries \textsc{SALSA-Mel}}\\
        
        \midrule

        SELDNet 
        & 5 & \textbf{0.632} & \textbf{0.276} & $\textbf{23.3}^\circ$ & 0.474 & \textbf{0.503} \\
        & 4 & 0.643 & 0.272 & $26.9^\circ$ & 0.462 & 0.515 \\
        & 3 & 0.665 & 0.245 & $28.8^\circ$ & \textbf{0.489} & 0.524 \\
        & 2 & \underline{0.680} & 0.251 & $30.0^\circ$ & 0.484 & 0.528 \\
        & 1 & 0.670 & \underline{0.218} & $\underline{30.1}^\circ$ & \underline{0.425} & \underline{0.549} \\

        \midrule

        ResNet-18 
        & 5 & \textbf{0.606} & \textbf{0.299} & $26.8^\circ$ & 0.466 & \textbf{0.497} \\
        & 4 & 0.613 & 0.270 & $\textbf{25.7}^\circ$ & 0.459 & 0.507 \\
        & 3 & 0.625 & 0.283 & $26.9^\circ$ & \textbf{0.486} & 0.501 \\
        & 2 & 0.633 & \underline{0.252} & $\underline{27.2}^\circ$ & 0.461 & 0.518 \\
        & 1 & \underline{0.648} & 0.254 & $27.0^\circ$ & \underline{0.458} & \underline{0.521} \\

        \midrule

        \multicolumn{2}{l}{\bfseries \textsc{MelGCC}}\\

        \midrule

        SELDNet 
        & 5 & \textbf{0.622} & \textbf{0.273} & $\textbf{23.7}^\circ$ & 0.463 & \textbf{0.504} \\
        & 4 & 0.632 & 0.266 & $27.9^\circ$ & \textbf{0.471} & 0.512 \\
        & 3 & 0.681 & 0.239 & $\underline{30.1}^\circ$ & 0.467 & 0.536 \\
        & 2 & \underline{0.690} & 0.237 & $27.8^\circ$ & 0.435 & 0.543 \\
        & 1 & 0.684 & \underline{0.201} & $28.8^\circ$ & \underline{0.386} & \underline{0.564} \\

        \midrule

        ResNet-18 
        & 5 & \textbf{0.602} & 0.272 & $27.3^\circ$ & 0.443 & \textbf{0.510} \\
        & 4 & 0.621 & \textbf{0.275} & $26.8^\circ$ & 0.441 & 0.514 \\
        & 3 & 0.643 & 0.274 & $27.0^\circ$ & \textbf{0.476} & 0.511 \\
        & 2 & 0.645 & \underline{0.245} & $\underline{27.7}^\circ$ & \underline{0.438} & \underline{0.522} \\
        & 1 & \underline{0.651} & 0.264 & $\textbf{26.5}^\circ$ & 0.445 & 0.516 \\

        \bottomrule

    \end{tabularx}
    \label{tab:input_segment_len}
\end{table}

As outlined in Section \ref{section:real_time_SELD_pipeline}, inference is performed on audio segments totaling $T_w$ seconds, with processing latency not exceeding $T_r$. The choice of $T_w$ directly impacts the size of input features and the resulting computational load of the inference module. While most SELD studies use larger $T_w$ values, up to 20 seconds for offline inference \cite{Du_NERCSLIP_task3_report}, this approach is generally impractical for real-time applications due to their high computational demands. For real-time inference, $T_w$ is typically shortened to minimize latency \cite{venkatesh_realtime_MSS}. As such, we first assess the impact of reduced $T_w$ durations on SELD performance. In particular, our experiments use two popular SELD model architectures -- SELDNet\cite{adavanne2018sound} and ResNet-18 \cite{kong2020panns}. For edge deployment, standard convolutions in ResNet-18 are replaced with Depthwise Separable Convolutions (DSCs)\cite{chollet2017xception} to reduce model complexity. 

Table \ref{tab:input_segment_len} summarizes the performance of various SELD systems as $T_w$ decreases. Results indicate that reducing $T_w$ generally leads to a decline in SELD performance. This performance degradation is likely due to the reduced temporal context in shorter audio segments, complicating the accurate identification and localization of sound events. Specifically, the SELDNet architecture exhibits an average degradation of 9.6\% in $\mathcal{E}_{\text{SELD}}$ when $T_w$ is reduced from 5 seconds to 1 second. In contrast, the ResNet-18 architecture shows only an average degradation of 3.3\% in $\mathcal{E}_{\text{SELD}}$ at the shortest $T_w$ tested, suggesting that more complex models are more robust against shorter input window durations. However, these performance gains come with increased computational cost and processing latency, highlighting a critical trade-off for real-time SELD.

\subsection{Real-time Inference Latency}
\label{section:real-time-inference}

\begin{table*}[ht!]
    \centering
    \caption{Computational costs of one inference pass using different model architectures and input features with $T_w = 2$. Dimensions refer to the number of feature channels $\times$ time bins $\times$ frequency or Mel bins. Processing latency is broken down into time required for feature extraction, model inference, and the remaining excess time before audio buffer overflow.}

    \label{tab:rpi_inference}
    \noindent\begin{tabularx}{0.82\textwidth}{lll cc rrr}
    \toprule
        Input Feature & Dimensions & System & Params (M) & MACs (G) & Feature (s) & Inference (s) & Excess (s) \\
        \midrule
        \bfseries \textsc{SALSA-Lite} 
        & $7 \times 160 \times 191$ 
        & SELDNet & 0.885 & 0.181 
        & 0.205 & 0.221 & 0.574 \\
        
        & 
        & ResNet-18 & 3.867 & 1.104 
        & 0.205 & 0.666 & 0.129 \\

        \midrule
        \bfseries \textsc{SALSA-Mel} 
        & $7 \times 160 \times 128$ & SELDNet & 0.835 & 0.125 
        & 0.434 & 0.169 & 0.397 \\
        
        & 
        & ResNet-18 & 3.867 & 0.774 
        & 0.434 & 0.463 & 0.103 \\

        \midrule
        \bfseries \textsc{MelGCC} 
        & $10 \times 160 \times 128$ & SELDNet & 0.837 & 0.161 
        & 0.433 & 0.198 & 0.369 \\
        
        & 
        & ResNet-18 & 3.868 & 0.791 
        & 0.433 & 0.488 & 0.079 \\

        \bottomrule
    \end{tabularx}
\end{table*}

The SELD models were deployed on a Raspberry Pi 3 to evaluate real-time inference capabilities. While methods such as quantization \cite{kamath2021performance_rpi} or pruning \cite{Bing2024_pruning} can reduce computational costs, these optimizations are not the focus of this study. In this work, we utilize the Open Neural Network Exchange (ONNX) inference engine for efficient execution on the Raspberry Pi 3 \cite{stefani2022comparison} and perform negligible post-processing (logging results). As such, for simplicity, the post-processing latency is combined with the inference latency. 

Table \ref{tab:rpi_inference} compares the performance of various SELD models using different input features, detailing parameter counts, multiply-and-accumulate (MACs) operations, and processing latencies. The parameter count determines the storage needed for model weights and MACs directly affect processing latency. The average processing latency is measured over 1000 iterations. The implementation in Table \ref{tab:rpi_inference} uses $T_r=1$, $n=2$, and $T_w=2$.

Table \ref{tab:rpi_inference} shows that the time needed for feature extraction can become significant, taking up to 43.4\% of $T_r$ in this context. As SELD tasks typically employ multi-channel recording devices \cite{Wilkins2023}, the increased number of audio channels amplifies the computational demand of feature extraction. The computational burden of feature extraction can be further analyzed through the algorithmic complexity of each process. Considering a single time frame, calculating Mel spectrograms from linear spectrograms incurs additional delay due to \eqref{eqn:melspec}, with complexity $\mathbf{O}(MF)$, where $F$ is the number of frequency points. NIPD calculation in SALSA-Lite rely on simple matrix multiplications of complexity $\mathbf{O}(MF)$, as compared to the more expensive GCC-PHAT computation performed for each unique microphone pair, or $(M^2 - M)/2$ times, resulting in a complexity of $\mathbf{O}(M^2F)$. Calculating SALSA-Mel from SALSA-Lite adds further latency due to the Mel filter dot product of complexity $\mathbf{O}(MF)$ in \eqref{eqn:salsa_mel}, causing its feature extraction time to be comparable to that of MelGCC.

\subsection{Design Considerations}

Real-time SELD systems need low latency, particularly for applications such as alarms where rapid response is critical. In this example pipeline, reducing $T_r$ lowers hardware latency, but imposes stricter constraints on processing latency, especially when maintaining $T_w$. Decreasing $T_w$ can help meet processing latency requirements, but this often reduces temporal context and degrades SELD performance, as shown in Table \ref{tab:input_segment_len}. More complex models can mitigate this performance loss, but at the cost of increased processing latency, creating a challenging trade-off consideration between model complexity, latency, and performance. 

\begin{figure}
    \centering
    \includegraphics[width=0.925\columnwidth]{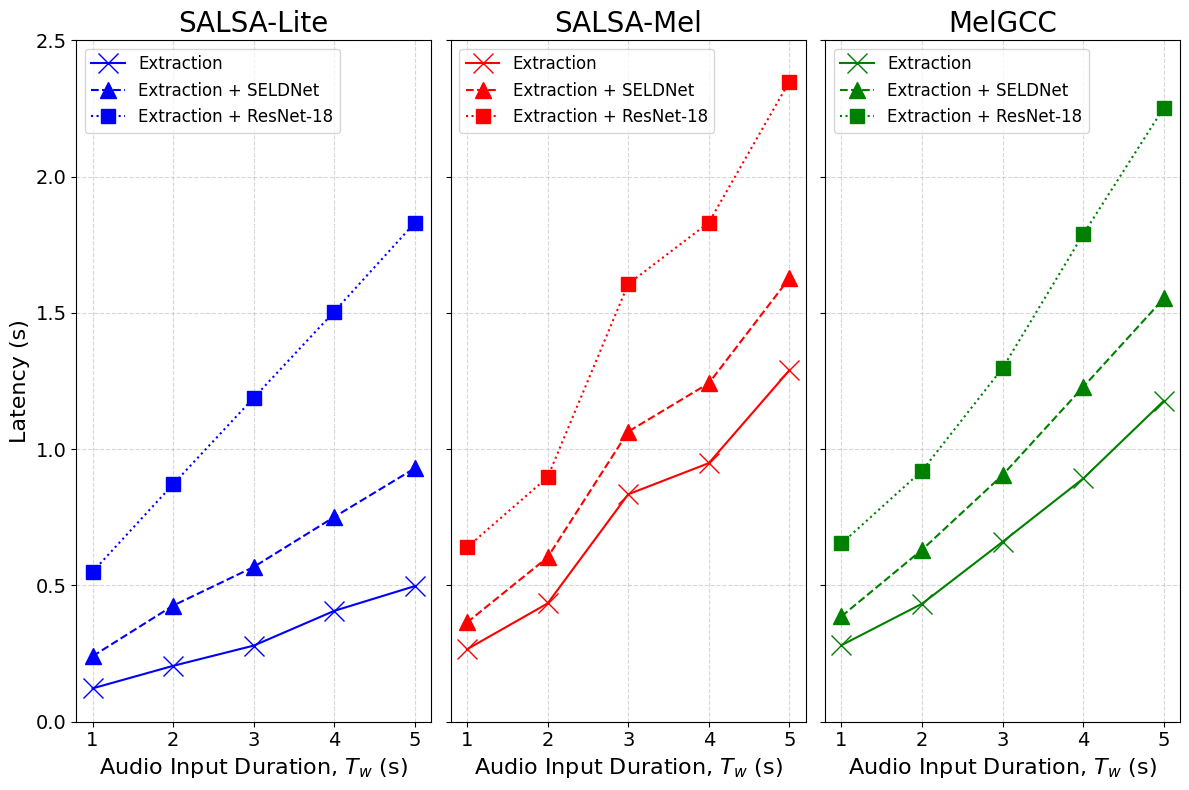}
    \caption{Breakdown of the average processing latency of different SELD systems against increasing $T_w$ over 1000 iterations.}
    \label{fig:feat_time_scaling}
\end{figure}

Table \ref{tab:rpi_inference} reveals another key point -- while reducing feature dimensionality slightly decreases MACs and processing latency, the computational overhead introduced by dimensionality reduction often negates these benefits. In contrast, computationally efficient features such as SALSA-Lite can instead offer significant reductions in overall latency, despite having larger feature dimensions. Figure \ref{fig:feat_time_scaling} illustrates how processing latency scales with increasing $T_w$, assuming fixed feature extraction parameters. The results suggest that relatively more computationally intensive features, such as MelGCC or SALSA-Mel, introduce substantial latency regardless of $T_w$. In contrast, the efficient SALSA-Lite brings about significant processing latency reductions with similar SELD performance.

\section{Conclusion}
\label{section:conclusion}

This paper highlights the critical challenges of deploying SELD systems for real-time applications on resource-constrained edge devices. Our experiments revealed that shorter input window durations, often necessary for low-latency inference, can degrade SELD performance by nearly $10\%$. This problem can be partially mitigated by adopting more complex DNN architectures, although at the cost of increased computational demand. Another key finding is that feature extraction constitutes a significant computational bottleneck, underscoring the need for optimized feature design. Balancing the trade-offs between performance, system complexity and latency will be crucial for advancing SELD systems for real-time inference. Future work advancing real-time SELD could involve moving receivers, or non-standard microphone configurations such as in smart wearable devices.

\newpage
\bibliographystyle{IEEEtran}
\bibliography{refs}

\end{document}